
%
\def\footnote#1{\let\@s\empty
  \ifhmode\edef\@s{\spacefactor=\the\spacefactor}\/\fi
  #1\@s\vfootnote{#1}}

\magnification=\magstep1
\baselineskip=20pt
\def\deltax{\delta_{q^\alpha,x}}
\def\deltay{\delta_{q^\beta,y}}
\def\n{\noindent}
\font\dai=cmr10 scaled \magstep2
\def\bigzero{\textfont0=\dai 0}
\def\m@th{\mathsurround=0pt}

\newdimen\LENB \newdimen\LENW \newdimen\THI
\newdimen\LENWH \newdimen\LENTOT \newcount\N
\def\vbrknlnele#1#2#3{
  \LENB=#1pt \LENW=#2pt \THI=#3pt
  \LENWH=\LENW \divide\LENWH by 2
  \LENTOT=\LENB \advance\LENTOT by \LENW
  \vbox to \LENTOT{
    \vbox to \LENWH{}
    \nointerlineskip
    \vbox to \LENB{\hbox to \THI{\vrule width \THI height \LENB}}
    \nointerlineskip
    \vbox to \LENWH{}
  }}

\def\vbrknln#1{
  \N=#1
  \vcenter{
    \vbox{
      \loop\ifnum\N>0
        \vbox to 4pt{\vbrknlnele{2}{2}{0.1}}
        \nointerlineskip
        \advance\N by -1
      \repeat
  }}}

\def\vbl#1{\hskip-5pt \vbrknln{#1} \hskip-5pt}

\def\hbrknlnele#1#2#3{
  \LENB=#1pt \LENW=#2pt \THI=#3pt
  \LENTOT=\LENB \advance\LENTOT by \LENW
  \vcenter{
    \vbox to \THI{
      \hbox to \LENTOT{
        \hfil
        \vrule width \LENB height \THI
        \hfil}
  }}}

\def\hblele{\hbrknlnele{2}{2.2}{0.1}}

\def\hblfil{\cleaders\hbox{$ \m@th \mkern1mu \hblele \mkern1mu $}\hfill}
\ \vskip20pt
\font\bigbf=cmb10 scaled\magstep2
\centerline{{\bigbf q-Discrete Toda Molecule Equation}}\par
\vskip18pt
\n Kenji Kajiwara\par
\n {\it Department of Applied Physics, Faculty of Engineering,}\par
\n {\it University of Tokyo, 7-3-1 Hongo, Bunkyo-ku, Tokyo 113, Japan}\par
\vskip18pt
\n Yasuhiro Ohta\footnote{$^1$}{On leave from
            Department of Applied Mathematics, Faculty of Engineering,
            Hiroshima University.}\par
\n {\it Research Institute for Mathematical Sciences, Kyoto University,}\par
\n {\it Kyoto 606, Japan}\par
\ \par
\n and \par
\ \par
\n Junkichi Satsuma\par
\n {\it Department of Mathematical Sciences, University of Tokyo,
3-8-1 Komaba, }\par
\n {\it Meguro-ku, Tokyo 153, Japan}\par
\vskip40pt
\noindent {\bf Abstract}\par
A q-discrete version of the two-dimensional Toda molecule equation
is proposed\par
\n through the direct method. Its solution, B\"acklund
transformation and Lax pair are discussed. The reduction to the q-discrete
cylindrical Toda molecule equation is also discussed. \par
\vfill\eject
\noindent {\bf 1. Introduction}\par
\vskip20pt
  The discovery of the quantum groups[1,2] gave rise to a new phase to
the studies of integrable systems. It also shed light to the theory
of so-called q-analog of the special functions, where the q-difference
operator plays a similar role to the differential operator in the
theory of the special functions[3-5]. On the other hand,
it is known that the soliton equations have a close relationship
with the special functions[6,7].  Hence it arises as an interesting
problem
to study ``q-discrete integrable systems" .\par
In a previous paper[8], we have proposed the q-discrete version
of the two-dimensional Toda lattice equation and shown that
the q-discrete cylindrical Toda lattice equation is derived
through a reduction. Moreover, we have shown that its solution is expressed
by a determinant whose entries are the q-Bessel function.
In this letter, we will discuss
the q-discretization of the two-dimensional Toda molecule (2DTM) equation
(finite, non-periodic two-dimensional Toda lattice)[9] and derive
its solution, B\"acklund transformation and Lax pair.
We also present the q-discrete cylindrical Toda molecule equation.\par
Let us first give a brief review of the 2DTM equation,

$$ {\displaystyle \partial^2 u_N\over\displaystyle \partial x\partial y}
={\rm e}^{u_{N-1}-u_N} - {\rm e}^{u_N - u_{N+1}}\ ,\quad N=1,\cdots,M
\ ,\eqno(1\rm a) $$
$$ u_0=-\infty,\qquad u_{M+1}=+\infty\ ,\eqno(1\rm b) $$

\n which is rewritten as

$$ \eqalignno{
{\displaystyle\partial V_N\over\displaystyle\partial x}
&=V_N( J_N-J_{N+1})\ ,&(2\rm a)\cr
{\displaystyle\partial J_N\over\displaystyle\partial y}
&=V_{N-1}-V_N\ ,&(2\rm b)\cr
V_0 &=V_M=0\ ,&(2\rm c)\cr} $$

\n by a suitable variable transformation. Equations (2a)-(2c) are
transformed to the bilinear form,

$${\displaystyle \partial^2\tau_N\over\displaystyle \partial x\partial y}
\tau_N - {\displaystyle \partial\tau_N\over\displaystyle \partial x}
{\displaystyle \partial\tau_N\over\displaystyle \partial y}
=\tau_{N+1}\tau_{N-1}\ , \eqno(3)$$

\noindent through the dependent variable transformations,

$$V_N = {\displaystyle \tau_{N-1}\tau_{N+1}\over\displaystyle \tau_N^2}\ ,
\qquad J_N = {\displaystyle \partial \over
\displaystyle \partial x}\ {\rm log}\ {\displaystyle \tau_{N-1}\over
\displaystyle \tau_{N}}\ .\eqno(4)$$

\n The solution of the bilinear form (3) is given by the two-directional
Wronskian,

$$\tau_N=\left\vert\matrix{
       f  &\partial_x f &\cdots &\partial_x^{N-1} f\cr
\partial_y f &\partial_x\partial_y f&\cdots &\partial_x^{N-1}\partial_y f
\cr
\vdots      &\vdots                &\ddots  &\vdots \cr
\partial_y^{N-1} f &\partial_x\partial_y^{N-1} f&\cdots &\partial_x^{N-1}
\partial_y^{N-1} f\cr}\right\vert\ . \eqno(5)$$

\n In (5), $f$ is chosen to satisfy the boundary condition (2c) as

$$f=\sum_{k=1}^M \phi_k(x)\chi_k(y)\ ,\eqno(6) $$

\n where $\phi_k(x)$ and $\chi_k(y)$ are arbitrary functions in
$x$ and $y$,
respectively.\par
Our method relies on the fact that the bilinear forms of
``generic" soliton equations are nothing but the identities of the
determinants or
the Pfaffians, or those derived from their reduction. In the present case,
(3) is reduced to the Jacobi identity of the two-directional Wronskian.
This fact can be a powerful
guiding principle for the extension of the integrable systems.
For example, usual discretization of soliton equations is performed
based on the principle[10]. We will apply the idea
to the q-discretization of the soliton equations. \par
\vskip20pt
\noindent {\bf 2. q-Discrete 2DTM Equation}\par
\vskip20pt
We propose a system given by
$$
\eqalignno{
\deltax V_N(x,y) &= V_N(q^\alpha x,y)J_N(x,q^\beta y) - V_N(x,y)
J_{N+1}(x,y)\ ,
&(7\rm a)\cr
\deltay J_N(x,y) &= V_{N-1}(q^\alpha x,y) - V_N(x,y)\ ,&(7\rm b)\cr
V_0(x,y)&=V_M(x,y)=0\ ,&(7\rm c)\cr }
$$

\n where $\deltax$ and $\deltay$ are the q-difference operators defined
by

$$\deltax f(x) = {\displaystyle f(x) - f(q^\alpha x)\over\displaystyle
(1-q)x}\ ,\qquad
\deltay f(y) = {\displaystyle f(y) - f(q^\beta y)\over\displaystyle
(1-q)y}\ .\eqno(8)$$

\n The operators $\deltax$ and $\deltay$ tend to
$\alpha {\displaystyle \partial \over\displaystyle \partial x}$ and
$\beta {\displaystyle \partial \over\displaystyle\partial y}$ in the limit
$q\rightarrow 1$, respectively. In this limit, (7a)-(7c)
are reduced to the 2DTM equation (2a)-(2c).
We call eqs. (7a)-(7c) the q-discrete
2DTM equation.\par

Equations (7a)-(7c) are transformed into the bilinear form,

$$\eqalignno{
& \deltax~\deltay \tau_N(x,y)\cdot \tau_N(x,y)
 - \deltax\tau_N(x,y)\deltay\tau_N(x,y)\cr
&\qquad = \tau_{N+1}(x,y)~\tau_{N-1}(q^\alpha x, q^\beta y)\ ,&(9)\cr} $$

\n through the dependent variable transformations,

$$\eqalignno{
J_N(x,y) &= {\displaystyle 1\over\displaystyle (1-q)x}\biggl\{
 {\displaystyle \tau_{N-1}(x,y)\tau_N(q^\alpha x,y)\over
\displaystyle \tau_{N-1}(q^\alpha x,y)\tau_N(x,y)} -1\biggr\}\ ,
&(10\rm a)\cr
V_N(x,y) &= {\displaystyle \tau_{N+1}(x,y)\tau_{N-1}(x,q^\beta y)\over
\displaystyle \tau_N(x,y)\tau_N(x,q^\beta y)}\ . &(10\rm b)\cr}$$

\n The solution of the bilinear form (9) is given by the
two-directional Wronski-type determinant,

$$
\tau_N(x,y)=\left\vert\matrix{
f(x,y) & \delta_{q^\alpha,x}f(x,y) &\cdots &\delta_{q^\alpha,x}^{N-1}
f(x,y)\cr
\delta_{q^\beta,y}f(x,y)&\delta_{q^\alpha,x}\delta_{q^\beta,y}f(x,y)
&\cdots &\delta_{q^\alpha,x}^{N-1}\delta_{q^\beta,y}f(x,y)\cr
\vdots  &\vdots &\ddots &\vdots \cr
\delta_{q^\beta,y}^{N-1}f(x,y)&\delta_{q^\alpha,x}\delta_{q^\beta,y}^{N-1}
f(x,y)
&\cdots &\delta_{q^\alpha,x}^{N-1}\delta_{q^\beta,y}^{N-1}f(x,y)\cr
}\right\vert \ ,\eqno(11)$$

\n where $f(x,y)$ is chosen to satisfy the boundary condition (7c) as (6).
\par
Let us prove that (11) really gives the solution of the bilinear form (9).
We have for example,

$$\eqalignno{
\tau_N(q^{-\alpha}x,y)
&=\left\vert\matrix{
f(q^{-\alpha}x,y) & \delta_{q^\alpha,x}f(q^{-\alpha}x,y) &\cdots
 &\delta_{q^\alpha,x}^{N-1}f(q^{-\alpha}x,y)\cr
\delta_{q^\beta,y}f(q^{-\alpha}x,y)
 &\delta_{q^\alpha,x}\delta_{q^\beta,y}f(q^{-\alpha}x,y)
 &\cdots &\delta_{q^\alpha,x}^{N-1}\delta_{q^\beta,y}f(q^{-\alpha}x,y)\cr
\vdots  &\vdots &\ddots &\vdots \cr
\delta_{q^\beta,y}^{N-1}f(q^{-\alpha}x,y)
 &\delta_{q^\alpha,x}\delta_{q^\beta,y}^{N-1}f(q^{-\alpha}x,y) &\cdots
 &\delta_{q^\alpha,x}^{N-1}\delta_{q^\beta,y}^{N-1}f(q^{-\alpha}x,y)\cr
}\right\vert\cr
&=\left\vert\matrix{
f(x,y) & \delta_{q^\alpha,x}f(x,y) &\cdots &\delta_{q^\alpha,x}^{N-1}
f(q^{-\alpha}x,y)\cr
\delta_{q^\beta,y}f(x,y)&\delta_{q^\alpha,x}\delta_{q^\beta,y}f(x,y)
&\cdots &\delta_{q^\alpha,x}^{N-1}\delta_{q^\beta,y}f(q^{-\alpha}x,y)\cr
\vdots  &\vdots &\ddots &\vdots \cr
\delta_{q^\beta,y}^{N-1}f(x,y)&\delta_{q^\alpha,x}\delta_{q^\beta,y}^{N-1}
f(x,y)&\cdots
&\delta_{q^\alpha,x}^{N-1}\delta_{q^\beta,y}^{N-1}f(q^{-\alpha}x,y)\cr
}\right\vert \ .&(12)}$$

\n In deriving the second determinant from the first,
we have subtracted $(k+1)$-th column multiplied by $(1-q)q^{-\alpha}x$
from $k$-th column for $k=1,\cdots N-1$, to confine the shift of the
independent variable $x$ to the most right column of the determinant.
Moreover, we note that $\deltax^{N-1}f(q^{-\alpha}x,y)$
means $\deltax^{N-1}f(x,y)\vert_{x\rightarrow q^{-\alpha}x}$.
Multiplying $N$-th column by $(1-q)q^{-\alpha}x$
and adding $(N-1)$-th column to $N$-th column in the second determinant of
(12), we get

$$\eqalignno{
&(1-q)q^{-\alpha}x~\tau_N(q^{-\alpha}x,y)\cr
&=\left\vert\matrix{
f(x,y) & \delta_{q^\alpha,x}f(x,y) &\cdots &\delta_{q^\alpha,x}^{N-2}
f(q^{-\alpha}x,y)\cr
\delta_{q^\beta,y}f(x,y)&\delta_{q^\alpha,x}\delta_{q^\beta,y}f(x,y)
&\cdots &\delta_{q^\alpha,x}^{N-2}\delta_{q^\beta,y}f(q^{-\alpha}x,y)\cr
\vdots  &\vdots &\ddots &\vdots \cr
\delta_{q^\beta,y}^{N-1}f(x,y)&\delta_{q^\alpha,x}\delta_{q^\beta,y}^{N-1}
f(x,y)&\cdots
&\delta_{q^\alpha,x}^{N-2}\delta_{q^\beta,y}^{N-1}f(q^{-\alpha}x,y)\cr
}\right\vert \ .&(13)}$$

\n Similarly, we obtain

$$\eqalignno{
&(1-q)^2q^{-(\alpha + \beta)}xy~\tau_N(q^{-\alpha}x,q^{-\beta}y)\cr
&=\left\vert\matrix{
 f(x,y) &\cdots &\deltax^{N-2}f(x,y) &\deltax^{N-2}f(q^{-\alpha}x,y)\cr
 \vdots &\ddots &\vdots &\vdots \cr
 \deltay^{N-2}f(x,y) &\cdots &\deltax^{N-2}\deltay^{N-2}f(x,y)
  &\deltax^{N-2}\deltay^{N-2}f(q^{-\alpha}x,y)\cr
 \deltay^{N-2}f(x,q^{-\beta}y) &\cdots
  &\deltax^{N-2}\deltay^{N-2}f(x,q^{-\beta}y)
  &\deltax^{N-2}\deltay^{N-2}f(q^{-\alpha}x,q^{-\beta}y)}\right\vert
 \ .&(14)}$$

\n Applying Jacobi's identity on (14) with $N$ replaced by $N+1$,
we obtain

$$ \eqalignno{
&\tau_N(q^{-\alpha}x,q^{-\beta} y)\tau_N(x,y) - \tau_N(q^{-\alpha}x,y)
\tau_N(x,q^{-\beta}y)\cr
& = (1-q)^2q^{-(\alpha + \beta)}xy~
\tau_{N+1}(q^{-\alpha}x,q^{-\beta} y)\tau_{N-1}(x,y)\ , &(15)\cr}$$

\n which is nothing but the bilinear form (9) with $x$ and $y$
replaced by $q^{-\alpha}x$ and $q^{-\beta}y$, respectively.
Thus we have proved that (11) gives the solution of (9).\par
We now discuss a reduction of the q-2DTM equation. Putting
$xy=r^2$ and $\alpha=\beta=2$, and imposing the condition that
$\tau_N(x,y)$ depends only on $r$,
we find that the bilinear form (9) and its solution (11)
are reduced to

$$
\bigl({\displaystyle 1\over\displaystyle r}\delta_{q,r}
+q\delta^2_{q,r}\bigr)\tau_N(r)\cdot\tau_N(r)-\bigl\{ \delta_{q,r}
\tau_N(r)\bigr\}^2
=\tau_{N+1}(r)\tau_{N-1}(q^2r)\ ,\eqno(16) $$

\n and

$$\eqalignno{
\tau_N(r) &= q^{ -{2\over 3}N(N-1)(N-2)}~r^{-N(N-1)}\cr
&\times\left\vert\matrix{
f(r) &r\delta_{q,r}f(r) &\cdots &(r\delta_{q,r})^{N-1}f(r)\cr
r\delta_{q,r}f(r) &(r\delta_{q,r})^2f(r) &\cdots &(r\delta_{q,r})^N
f(r)\cr
\vdots &\vdots &\ddots &\vdots \cr
(r\delta_{q,r})^{N-1}f(r) &(r\delta_{q,r})^Nf(r) &\cdots
&(r\delta_{q,r})^{2N-2}f(r)\cr}\right\vert \ ,&(17)\cr}$$

\n respectively. Equation (16) tends to the cylindrical Toda molecule
(cTM) equation[11],

$$
\bigl({\displaystyle 1\over\displaystyle r}{\displaystyle \partial
\over\partial r}
+{\displaystyle\partial^2\over\displaystyle \partial r^2}\bigr)\tau_N(r)
\cdot\tau_N(r)-\bigl\{ {\displaystyle\partial \tau_N(r)\over\displaystyle
\partial r}\bigr\}^2=\tau_{N+1}(r)\tau_{N-1}(r)\ ,\eqno(18) $$

\n in the limit $q\rightarrow 1$, and hence we call (16) the
q-cTM equation. Note that (16) is transformed to

$$\eqalignno{
& \delta_{q,r}V_N(r) = qJ_N(qr)V_N(qr) - J_{N+1}(r)V_N(r)\ ,&(19\rm a)\cr
& \bigl( q\delta_{q,r}+{\displaystyle 1\over\displaystyle r}\bigr)
J_N(r)=V_N(r) - V_{N-1}(qr)\ ,&(19\rm b)\cr
& V_0(r) = V_M(r)=0\ ,&(19\rm c)\cr}
$$

\n through the dependent variable transformations,

$$ \eqalignno{
V_N(r) &= {\displaystyle \tau_{N-1}(qr)\tau_{N+1}(r)\over\tau_N(r)
\tau_N(qr)}\ ,&(20\rm a)\cr
J_N(r) &={\displaystyle 1\over\displaystyle (1-q)r}\biggl\{
{\displaystyle \tau_N(qr)\tau_{N-1}(r)\over\tau_N(r)\tau_{N-1}(qr)}
- 1\biggr\}\ .&(20\rm b)\cr}$$
\vskip20pt


\n {\bf 3. B\"acklund Transformation and Lax Pair}\par
\vskip20pt

By using the fact that the solution of the q-2DTM equation (7) is
given by (11), we here propose the B\"acklund transformation. It is
written by

$$\eqalignno{
&\deltay~\tau_N(x,y)\cdot\tau_N^\prime (x,y) - \tau_N(x,y)~\deltay~
\tau_N^\prime (x,y)\cr
&\qquad =- \tau_{N+1}(x,y)\tau_{N-1}^\prime (x,q^\beta y)\ ,
&(21\rm a)\cr
&\deltax~\tau_N(x,y)\cdot\tau_{N-1}^\prime (x,y) - \tau_N(x,y)~\deltax~
\tau_{N-1}^\prime (x,y)\cr
&\qquad =\tau_{N-1}(q^\alpha x,y)\tau_N^\prime (x,y)\ ,&(21\rm b)\cr} $$

\n which transforms a solution of the q-2DTM equation,

$$
\tau_N(x,y)=\left\vert\matrix{
f(x,y) & \delta_{q^\alpha,x}f(x,y) &\cdots &\delta_{q^\alpha,x}^{N-1}
f(x,y)\cr
\delta_{q^\beta,y}f(x,y)&\delta_{q^\alpha,x}\delta_{q^\beta,y}f(x,y)
&\cdots &\delta_{q^\alpha,x}^{N-1}\delta_{q^\beta,y}f(x,y)\cr
\vdots  &\vdots &\ddots &\vdots \cr
\delta_{q^\beta,y}^{N-1}f(x,y)&\delta_{q^\alpha,x
}\delta_{q^\beta,y}^{N-1}f(x,y)
&\cdots &\delta_{q^\alpha,x}^{N-1}\delta_{q^\beta,y}^{N-1}f(x,y)\cr
}\right\vert\ ,\eqno(22)$$

\n to another solution,

$$
\tau_N^\prime (x,y)=\left\vert\matrix{
\deltax f(x,y) & \delta_{q^\alpha,x}^2f(x,y) &\cdots
&\delta_{q^\alpha,x}^{N}f(x,y)\cr
\deltax\delta_{q^\beta,y} f(x,y)&\delta_{q^\alpha,x}^2
\delta_{q^\beta,y}f(x,y)
&\cdots &\delta_{q^\alpha,x}^{N}\delta_{q^\beta,y}f(x,y)\cr
\vdots  &\vdots &\ddots &\vdots \cr
\deltax \delta_{q^\beta,y}^{N-1}f(x,y)&\delta_{q^\alpha,x}^2
\delta_{q^\beta,y}^{N-1}f(x,y)
&\cdots &\delta_{q^\alpha,x}^{N}\delta_{q^\beta,y}^{N-1}f(x,y)\cr
}\right\vert\ . \eqno(23)$$

\n In other words, (21a) and (21b) are the identities
for the determinants (22) and (23).
This fact is shown by the Pl\"ucker relation as follows. \par
Let us prove the
second equation (21b).  First, we introduce notations

$$\tau_N(x,y)= \vert 0, 1, \cdots , N-1\vert\ , \eqno(24)$$
$$\tau_N^\prime (x,y)= \vert 1, 2, \cdots , N\vert\ . \eqno(25)$$

\n Namely, the number $``k"$ in (24) and (25) means a column vector

$$``k"=\left.\pmatrix{ \deltax^k f(x,y)\cr
                \deltax^k\deltay f(x,y)\cr
                \vdots\cr
                \deltax^k\deltay^{N-1}f(x,y)\cr}\right\}{\scriptstyle N}
\ . \eqno(26)$$

\n Then we have

$$\tau_{N-1}(x,y) = \vert 0,1,\cdots , N-2, \phi\vert\ ,\eqno(27)$$

$$\tau_N(q^{-\alpha}x, y) = \vert 0, 1, \cdots , N-2,
N-1_{q^{-\alpha}x}\vert\ ,\eqno(28)$$

$$\tau_{N-1}^\prime(q^{-\alpha}x,y)=\vert 1, 2 , \cdots ,
N-2, N-1_{q^{-\alpha}x}, \phi \vert\ , \eqno(29)$$

\n and

$$(1-q)q^{-\alpha}x\ \tau^\prime_N(q^{-\alpha}x,y) =
\vert 1, 2,\cdots N-1, N-1_{q^{-\alpha}x}\vert\ , \eqno(30)$$

\n where

$$``N-1_{q^{-\alpha}x}"=\pmatrix{ \deltax^{N-1} f(q^{-\alpha}x,y)\cr
                     \deltax^{N-1}\deltay f(q^{-\alpha}x,y)\cr
                     \vdots\cr
                     \deltax^{N-1}\deltay^{N-1}f(q^{-\alpha}x,y)\cr}\ ,
\eqno(31)$$

\n and

$$\phi = \left.\pmatrix{
                        0\cr
                        \vdots\cr
                        0\cr
                        1\cr}\right\} {\scriptstyle N}\ , \eqno(32)$$

\n which is inserted to equalize the size of the determinant.
\par

We now consider an identity of $2N\times 2N$ determinant,

$$0=\left\vert\matrix{
0&\vbl4&1 &\cdots &N-2&\vbl4 &\ &\bigzero &\ &\vbl4 &N-1
&N-1_{q^{-\alpha}x}&\phi\cr
\multispan {13}\hblfil\cr
0&\vbl4&\ &\bigzero&\ &\vbl4 &1 &\cdots &N-2 &\vbl4 &N-1
&N-1_{q^{-\alpha}x}&\phi\cr
}\right\vert\ .\eqno(33)$$

\n Applying the Laplace expansion to the right-hand side, we obtain
an identity ( the Pl\"ucker relation),

$$\eqalignno{
0=&\vert 0,1,\cdots, N-2, N-1_{q^{-\alpha}x}\vert\
 \vert 1,\cdots, N-2, N-1, \phi\vert\cr
-&\vert 0,1,\cdots, N-2, N-1\vert\
 \vert 1,\cdots, N-2, N-1_{q^{-\alpha}x},\phi\vert\cr
-&\vert 1,\cdots, N-2, N-1,N-1_{q^{-\alpha}x}\vert\
 \vert 0,1,\cdots, N-2, \phi\vert\ ,&(34)\cr}$$

\n or equivalently,

$$\eqalignno{
&\tau_N(q^{-\alpha}x,y)\tau_{N-1}^\prime (x,y) - \tau_N(x,y)
\tau_{N-1}^\prime (q^{-\alpha}x,y)\cr
&\qquad = (1-q)q^{-\alpha}x~\tau_{N-1}(x,y)
\tau_N^\prime (q^{-\alpha}x,y)\ ,&(35)\cr}$$

\n which is nothing but (21b) with $x$ replaced by
$q^{-\alpha}x$. Thus we have completed the proof. The first equation (21a)
is proved in a similar way. \par

It is possible in general to construct the Lax pair from
the B\"acklund transformation.
Following the method developed by Hirota {\it et.al.}[12], we derive
the Lax pair for the q-2DTM equation (7) from (21). Introducing
$\psi$ by

$$ \tau_N^\prime (x,y)=\tau_N(x,y)\psi_{N+1}(x,y)\ ,\eqno(36)$$

\n we have from (21),

$$\eqalignno{
\deltay \psi_{N+1}(x,y) &= V_N(x,y)\psi_N(x,q^\beta y)\ ,&(37\rm a)\cr
\deltax \psi_N(x,y) &=-J_N(x,y)\psi_N(x,y) - \psi_{N+1}(x,y)\ .&(37\rm b)
\cr} $$

\n Let us define two matrices $L$ and $R$ by

$$L(x,y) = \pmatrix{
                   0\ \     &\       &\       &\        \cr
                   V_1(x,y) &0\ \ \  &\bigzero&         \cr
                   \        &\hskip-20pt\ddots  &\hskip-20pt \ddots &\
\cr
                   \bigzero &\       &\       &\        \cr
                   \        &\       &\hskip-20pt V_{M-1}(x,y)&0    \cr}
\ ,\eqno(38)$$

$$R(x,y) = - \pmatrix{
      J_1(x,y) &1       &\       &\        &\ \cr
      \        &\hskip-20pt J_2(x,y)&1 \ \ \  &\bigzero &\ \cr
      \        &\       &\ddots  &\ddots   &\ \cr
      \bigzero &\       &\       &\hskip-30pt J_{M-1}(x,y) &1 \cr
      \        &\       &\       &\            &\hskip-20pt J_M(x,y)
\cr}\ .\eqno(39)$$

\n Then (37) are rewritten as

$$\deltay\Psi(x,y) = L(x,y)\Psi(x,q^\beta y)\ ,
\qquad \deltax\Psi(x,y) = R(x,y)\Psi(x,y)\ ,\eqno(40)$$

\n where

$$\Psi(x,y)=\pmatrix{ \psi_1(x,y) \cr
                      \vdots \cr
                      \psi_M(x,y)\cr}\ . \eqno(41)$$

\n The compatibility condition of the linear system (40) yields

$$\deltax L(x,y) - \deltay R(x,y) = R(x,y)L(x,y)
- L(q^\alpha x,y)R(x,q^\beta y)\ , \eqno(42)$$

\n which recovers the q-2DTM equation (7). Consequently,
(38) and (39) give the Lax pair of the q-2DTM equation. \par
\vskip20pt
\n {\bf 4. Concluding Remarks}\par
\vskip20pt

In this letter, we have proposed the q-2DTM equation,
 its solution, B\"acklund transformation and Lax pair.
There are several definitions of integrability for the continuous
equations, such as the existence of N-soliton solution,
that of an infinite number of conserved quantities or symmetries,
Painlev\'e property and so on. For a given continuous equation,
there are several ways to
discretize it, depending on the definition to be taken.
In our case, the guiding principle of discretization is to
preserve integrability in such a sense that the determinant
structure of solutions of the discretized equation
is the same as that of the original continuous equation.
For the discretization we have employed the bilinear formalism,
since the determinant structure is clearly seen in it.\par
It was revealed that the time evolution of the solutions of
the usual integrable system is subject to
the infinite dimensional Lie algebra[13]. A question naturally
arising is
what the algebra describing the q-discrete system is.
Is it the quantum group or some others?
It may be an interesting and important problem
to find the structure of the algebras underlying the q-2DTM equation.\par
Finally, we mention the possibility of q-discretization
of the Painlev\'e equations. There is a close
relationship between the Painlev\'e equations and
Toda molecule equation.\break
Okamoto has shown that
the $\tau$ functions of Painlev\'e equations
satisfy the several types of Toda molecule equations[14].
In particular, the $\tau$ function of
the third Painlev\'e equation satisfies
the cTM equation. Moreover, it has been shown that
the Painlev\'e equations are reduced to
the identities of determinants[15,16] for special values
of the parameters in the equation.
Since we have  the q-cTM equation,
it is natural to expect
that we can perform the q-discretization of the third Painlev\'e
equation. We will report on this subject in a forthcoming paper.\par

\vskip20pt
\n {\bf References}\par
\vskip20pt
\item{[1]} V.~G.~Drinfel'd, Sov.~Math.~Dokl. {\bf 32}(1985) 254.
\item{[2]} M.~Jimbo, Lett.~Math.~Phys. {\bf 10}(1985) 63.
\item{[3]} H.~Exton, q-Hypergeometric Functions and Applications
(Ellis Horwood, Chi-\break chester, 1983).
\item{[4]} T.~H.~Koornwinder, in: Orthogonal Polynomials ed. P. Nevai
(Kluwer Academic, Dordrecht, 1990) p.257.
\item{[5]} T.~Masuda, K.~Mimachi, Y.~Nakagami, M.~Noumi and K.~Ueno,
J.~Func.~Anal. {\bf 99} (1991) 357.
\item{[6]} A.~Nakamura, Prog.~Theor.~Phys.~Suppl. {\bf 94} (1988)195.
\item{[7]} K.~Okamoto, in: Algebraic Analysis (Academic, Boston, 1988)
p.647.
\item{[8]} K.~Kajiwara and J.~Satsuma, J.~Phys.~Soc.~Jpn. {\bf 60}(1991)
3986.
\item{[9]} A.~N.~Leznov and M.~V.~Saveliev, Physica 3D (1981) 62.
\item{[10]} Y.~Ohta, R.~Hirota, S.~Tsujimoto and T.~Imai, submitted to
J.~Phys.~Soc.~Jpn.
\item{[11]} R.~Hirota and A.~Nakamura, J.~Phys.~Soc.~Jpn. {\bf 56}(1987)
3055.
\item{[12]} R.~Hirota, S.~Tsujimoto and T.~Imai, to appear in Proc. of
Future Directions of Nonlinear Dynamics in Physics and Biological Systems
(1992).
\item{[13]} M.~Jimbo and T.~Miwa, Publ.~RIMS, Kyoto Univ. {\bf 19}(1983)
 943.
\item{[14]} K.~Okamoto, Ann.~Mat.~Pura Appl. {\bf 146}(1987) 337;
Japan J. Math. {\bf 13}(1987) 47; Math. Ann. {\bf 275}(1986) 221;
Funkcial.~Ekvac. {\bf 30}(1987) 305.
\item{[15]} A.~Nakamura, J.~Phys.~Soc.~Jpn. {\bf 61}(1992) 3007.
\item{[16]} K.~Kajiwara, Y.~Ohta and J.~Satsuma, in preparation.

\bye